\documentclass[journal,comsoc]{IEEEtran}

\usepackage{cite}

\ifCLASSINFOpdf
  \usepackage[pdftex]{graphicx}
\else
  \usepackage[dvips]{graphicx}
\fi

\usepackage{amsmath}

\ifCLASSOPTIONcompsoc
 \usepackage[caption=false,font=normalsize,labelfont=sf,textfont=sf]{subfig}
\else
 \usepackage[caption=false,font=footnotesize]{subfig}
\fi

\usepackage{url}

\usepackage{comment}
\usepackage{listings}
\usepackage{listings}
\usepackage{xcolor}
\usepackage{booktabs}
\usepackage{orcidlink}

\begin{document}
%
\title{Mitigating IoT Botnet DDos Attacks through MUD and eBPF based Traffic Filtering}
%
%
%

\author{Angelo~Feraudo\orcidlink{0000-0002-9727-0861},
        Diana~Andreea~Popescu\orcidlink{0000-0002-2435-9603},
        Poonam~Yadav\orcidlink{0000-0003-0169-0704},
        Richard~Mortier\orcidlink{0000-0001-5205-5992}, and
        Paolo~Bellavista\orcidlink{0000-0003-0992-7948} 
\thanks{A. Feraudo and P. Bellavista are with the Department
of Computer Science and Engineering, University of Bologna, Bologna, 40136
Italy (e-mail: name.surname@unibo.it)}
\thanks{D. A. Popescu and R. Mortier are with the Department of Computer Science and Technology, University of Cambridge, Cambridge, CB3 0FD UK (email: diana.popescu@cl.cam.ac.uk and richard.mortier@cl.cam.ac.uk)}
\thanks{P. Yadav is with the Department of Computer Science, University of York, York, YO10 5GH UK (email:poonam.yadav@york.ac.uk)}
}

%
%

\markboth{In review, May~2023}%
{Feraudo \MakeLowercase{\textit{et al.}}: }
%



\maketitle

\begin{abstract}
As the prevalence of Internet-of-Things (IoT) devices becomes more and more dominant, so too do the associated management and security challenges. One such challenge is the exploitation of vulnerable devices for recruitment into botnets, which can be used to carry out Distributed Denial-of-Service (DDoS) attacks. The recent Manufacturer Usage Description (MUD) standard has been proposed as a way to mitigate this problem, by allowing manufacturers to define communication patterns that are permitted for their IoT devices, with enforcement at the gateway home router. In this paper, we present a novel integrated system implementation that uses a MUD manager (osMUD) to parse an extended set of MUD rules, which also allow for rate-limiting of traffic and for setting appropriate thresholds. Additionally, we present two new backends for MUD rule enforcement, one based on \emph{eBPF} and the other based on the Linux standard \emph{iptables}. The reported evaluation results show that these techniques are feasible and effective in protecting against attacks and in terms of their impact on legitimate traffic and on the home gateway. 

  
\end{abstract}


\newcommand\note[2]{\color{#1}\bf #2}
\newcommand\mort[1]{{\note{red}{mort: #1}}}

\newcommand{\one}{({\em i})\/}
\newcommand{\two}{({\em ii})\/}
\newcommand{\three}{({\em iii})\/}
\newcommand{\four}{({\em iv})\/}
\newcommand{\five}{({\em v})\/}

\newcommand\s[1]{(\S\ref{#1})}

%
\IEEEpeerreviewmaketitle

\section{Introduction}

\IEEEPARstart{A}{s} the use of Internet-of-Things (IoT) devices, in particular in homes and in open deployment environments, continues to increase, numerous associated security challenges have emerged. The particular challenge we address in this work is the hijacking of IoT devices for recruitment into botnets. Such devices are Internet-connected by design, typically through high-bandwidth home broadband connections, and are often not directly and actively used by residents. In the hijacking process of IoT devices, attackers exploit vulnerabilities in the security of the device, gain unauthorized access, and take control of the device. Once the device is under the attacker's control, it can be used to perform a range of malicious activities, such as generating spam and stealing sensitive information.

One common method used by bots to target IoT devices is through brute-force attacks on default or weak login credentials. This enables attackers to gain access to the device and take control of it. Furthermore, IoT devices that are not regularly updated may have known vulnerabilities that attackers can exploit to gain control. As a consequence, once an IoT device is hacked and recruited into a botnet, its intended function will continue to work, but it can also generate massive amounts of network traffic as part of a Distributed Denial-of-Service (DDoS) attack.

Ideally, IoT devices would not be as susceptible to hacking, but as long as they remain vulnerable, one approach to detecting and mitigating the effects of hacked devices is to monitor and control the traffic they generate. The IETF has attempted to address this issue through RFC 8520, which is the ``Manufacturer Usage Description Specification"~\cite{rfc8520, Krishnan2022}. This specification allows IoT device manufacturers or service providers to provide a machine-readable description of the network interaction in which the device should engage. This information can be used by the network home router to enforce restrictions on devices' network activities. Consequently, a hacked device cannot be effectively used as part of a botnet because it will not be allowed to generate arbitrary Internet traffic.

Specifically, these MUD files allow manufacturers or service providers to specify to the local network all the permitted incoming and outgoing source/destination addresses, enabling fine-grained traffic filtering of each IoT device. However, it does not allow any further constraint, for example on the rate or mix of traffic generation. Further, implementing the specified constraints on relatively resource-limited home routers requires efficient means to intercept and control network traffic as the number of constraints rises in proportion to the number of IoT device types in the home. Thus, the MUD purpose is to restrict traffic while enabling IoT manufacturers to provide long-term support and updates for their devices, supporting and facilitating the scalable deployment and management of IoT ecosystems. 

This paper advances the state-of-the-art of the related work in the field by proposing the following original contributions:

\begin{enumerate}
\item we propose extensions to the existing MUD standard that enable fine-grained rate-limiting of traffic from controlled devices, alongside means to estimate the relevant parameter values in realistic deployments~\s{s:extending-mud};
\item we extend the existing osMUD~\cite{osMud} implementation so that it can be deployed within a virtual machine setup for development and testing or within a real-world setup~\s{s:liberating-mud}; and
\item we develop and integrate new backends for osMUD by using Linux standard \emph{iptables} firewalls and eBPF~\cite{ebpf} that support more efficient implementation of these constraints as expressed in MUD files~\s{s:adapting-mud}.
\end{enumerate}

In addition to the above original contributions (to the best of our knowledge, the presented prototype is the first implementation proposed for the employment of MUDs to mitigate IoT DDoS attacks), the paper contributes to the literature in the field by reporting novel quantitative performance results on how our design/implementation choices permit to limit high traffic surges caused by a Botnet attack~\cite{kolias2017ddos}, demonstrating their effectiveness~\s{s:evaluation}.

\section{Extending MUD}
\label{s:extending-mud}

According to the MUD standard~\cite{rfc8520}, a MUD deployment consists of three architectural building blocks: \one~the device behaviour description (MUD file), \two~a uniform resource locator (MUD URL) and \three~a mechanism for local management systems that uses the URLs to request description files. In addition, the standard defines two main components that guarantee deployment and use of MUD files: the MUD file server that makes description files available, and the MUD manager, which requests and receives description files to and from the MUD file server.

The workflow between these blocks requires the thing or device to emit the MUD URL indicating where the corresponding MUD file is hosted. For this purpose, three protocol extensions have been defined by the standard: \one~in DHCP, a reserved option in request packets is used; \two~in X.509 through a certificate extension; and \three~in Link Layer Discovery Protocol (LLDP) by exploiting a subtype defined in RFC 7042. Once a MUD file has been retrieved, the MUD manager validates and enforces the Access Control Lists (ACLs) produced on the corresponding firewall.

\subsection{The MUD data model}

The MUD data model consists of a YANG based file serialised in JSON~\cite{rfc7951}. The YANG language provides a simple way to model different types of data, such as configuration data and notifications for network management protocols. There are only three YANG schema components that are serialised in a MUD file: \one~\emph{ietf-mud} allows to verify MUD file validity as well as the policy to and from the device; \two~\emph{ietf-access-control-list}~\cite{rfc8519} defines Access Control Lists by using a YANG data model; \three~\emph{ietf-acldns} allows the DNS matching criteria.

The ACL model feature involves the main network and transport layer protocols (\emph{ipv4}, \emph{ipv6}, \emph{udp}, \emph{tcp}, \emph{icmp}). The actions included in this model are \emph{ACCEPT} or \emph{DROP}, the \emph{REJECT} action can be interpreted by the MUD Manager as \emph{DROP}. For the osMUD manager, \emph{REJECT} is the default action to deny device communications towards any kind of domain not included in its MUD file. Hence, the MUD file model allows Manufacturers to define a set of white lists that describe the services needed by their devices to work properly.

\subsection{A rate-limiting extension}
\label{ss:extending-mud-rate-limiting}

The MUD data model extension requires inclusion of a new field in the MUD file related to rate-limits. The MUD file contains Access Control Lists (ACLs) as defined in IETF RFC 8519~\cite{rfc8519}. Each of them comprises a list of rules known as Access Control Entries (ACEs), which correspond to a group of match criteria and a group of actions. According to the standard, a rate-limit operation belongs to the action group of an ACE. Therefore, we add rate-limits defined by \emph{packet-rate} and \emph{byte-rate} along with the \emph{forwarding} field in the \emph{actions} object, as shown in Fig.~\ref{lst:actions}.

\begin{figure}
  \begin{lstlisting}
    "actions": {
      "packet-rate": "50/second",
      "byte-rate": "50kb/minute",
      "forwarding": "accept"
    }
  \end{lstlisting}
  \caption{The \emph{actions} object after rate-limits addition.
    \label{lst:actions}}
\end{figure}

Manufacturers may use these fields to define rate-limits using different metrics, such as second, minute, hour, and day. As these fields belong to the action group, manufacturers can set them for each allowed communication enabling a more precise and targeted approach. Furthermore, as shown in Fig.~\ref{lst:actions}, they do not necessarily have to be the same in both packet- and byte-rates. We discuss the implementation of our extension in Section~\ref{s:liberating-mud}.



\subsection{Learning thresholds}

To tune our proposed extension for real deployment environments, we must extract information to produce valid upper bounds for outgoing traffic. We have performed this analysis on the widely accepted data gathered by Ren et al.~\cite{ren2019information}. This includes different PCAP files containing network traffic generated by $81$ IoT devices of six different categories: \emph{appliances}, \emph{smart-hubs}, \emph{cameras}, \emph{audio} \emph{home-automation}, and \emph{televisions}. Those PCAP files were processed to compute the amounts of packets and bytes sent and received by each device in a given time window, defined as follows: the first packet timestamp defines the window start time ($wst$); if the packet falls within the window ($wst + windowSize$), the packet/byte counters corresponding to this window are incremented; otherwise, while that packet cannot be included within a window, the next window start time is updated with the end time of the previous one ($wst_{t}\gets wst_{t-1} + window\_size$). We analyzed the data using a window of $60$ seconds.

For the sake of brevity, in the paper we define and validate MUD files for two device categories: \emph{appliances} and \emph{smart-hubs}. The former includes devices that assist home activities, such as cooking, cleaning and printing. The latter involves access points or controllers that provide Internet access to IoT devices, whose communications may take place via either proprietary or low-range protocols. Moreover, considering our initial assumptions, i.e., preventing Botnet attacks generated from IoT devices, the analysis reports only data related to devices' outgoing traffic.

\begin{table}
  \caption{Appliances and smart-hubs outgoing traffic in 60 seconds
    \label{table:data}}
  \renewcommand{\arraystretch}{1.2}
  \begin{tabular}{ l r r r r }
    \toprule
    Category           & TCP     & TCP Max   & UDP     & UDP Max \\
    \midrule
    appliances (pkts)  & 36.7    & 223.2      & 5.033   & 140.28\\
    smart-hubs (pkts)  & 21.3    & 1716.8     & 9.63    & 152.38\\
    cameras (pkts)     & 62.2    & 1471.34    & 94.40   & 7863.0\\
    audio (pkts)       & 52.01   & 6687.5     & 293.6   & 1837.63\\
    home-aut (pkts)    & 5.20    & 702.3      & 14.51   & 127.5\\
    tv (pkts)          & 128.9   & 7560       & 33.33   & 729.3\\
    \midrule
    appliances (bytes) & 2350.7  & 36761.2    & 1446    & 24390.2\\
    smart-hubs (bytes) & 2375.3  & 177385.12  & 1970    & 38507.9\\
    cameras (bytes)    & 75065   & 1257726.67 & 81198   & 6727372.3\\
    audio (bytes)      & 14619.7 & 3430505.1  & 18758.1 & 141291.4\\
    home-aut (bytes)   & 1084.5  & 56560.5    & 3224.63 & 24966.8\\
    tv (bytes)         & 27525.7 & 2503001    & 6734.5  & 80519.3\\
    \bottomrule
  \end{tabular}
\end{table}

Table~\ref{table:data} summarises the results of our analysis in terms of devices' outgoing traffic in a window of $60$ seconds. The \emph{TCP} and \emph{UDP} columns refer to the average of TCP and UDP packets/bytes outgoing traffic during device activities, thus do not consider idle periods (e.g.,~during the night). Similarly, the other columns indicate the network traffic peaks average for each category. The data show that devices in both considered categories, \emph{appliances} and \emph{smart-hubs}, require TCP to work properly.

We define two MUD files for each device category analyzed, describing allowed reliable communications and including two volumetric policies, namely ``peaks'' and ``averages''. Hence, looking at Table~\ref{table:data}, MUD files using ``peaks'' as outgoing traffic threshold define 250 packets per minute and 40\,kB per minute for devices belonging to \emph{appliances} category, and 1720 packets per minute and 180\,kB per minute for \emph{smart-hubs} devices. On the other hand, MUD files using ``average'' outgoing traffic during device activities define 40 packets per minute and 3\,kB per minute for \emph{appliances}, and 22 packets per minute and 3\,kB per minute for \emph{smart-hubs}. It is worth noting that selected limits concern all the devices in the categories analysed. However, as described in section \ref{ss:extending-mud-rate-limiting}, the proposed MUD data model allows manufacturers to define multiple limits for each device, i.e., one for each allowed destination. Defining a limit per destination will create fine-grained policies for each device, preventing
DoS attacks originating from devices whose category aggregate rate limit is higher than their peak rate.

We leverage this analysis to establish the rate limits that we use to validate the effectiveness of our MUD model proposal for both normal and abnormal IoT network traffic.

\section{Liberating MUD}
\label{s:liberating-mud}

\begin{figure}
  \centering
  \includegraphics[width=\linewidth]{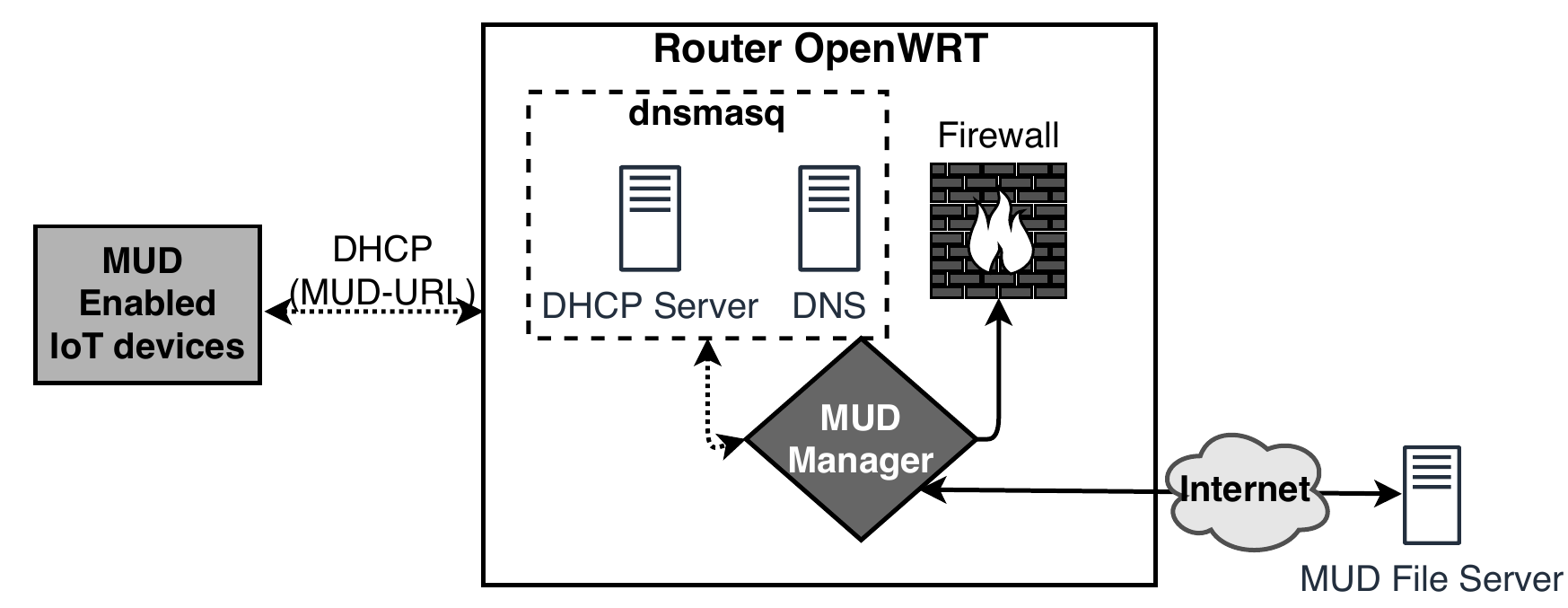}
  \caption{osMUD Architecture
    \label{fig:osMUDArchitecture}}
\end{figure}

Open Source MUD (osMUD)~\cite{osMud} is an open-source implementation of MUD manager, developed by a consortium of device manufacturing and network security companies. As shown in Figure~\ref{fig:osMUDArchitecture}, the MUD manager is designed to be easily built, deployed, and run on Open Wireless Router (OpenWRT) platform. The implementation requires a customised version of dnsmasq to enable MUD URL extraction, provide network infrastructure services, and minimise resource usage.

From the MUD data model perspective, the current implementation does not support MUD file rules for lateral movement attacks (e.g.,~same-manufacturer, controller, my-controller)~\cite{rfc8520}. Thus, after compromising a device inside the MUD compliant network, adversaries can progressively move through other systems, searching for targeted key data and assets, which are exploited to gain access to other hosts or applications within the network.  

\begin{figure*}
  \centering
  \subfloat[Output flow\label{fig:networkdep-out}]{%
    \includegraphics[width=0.48\linewidth]{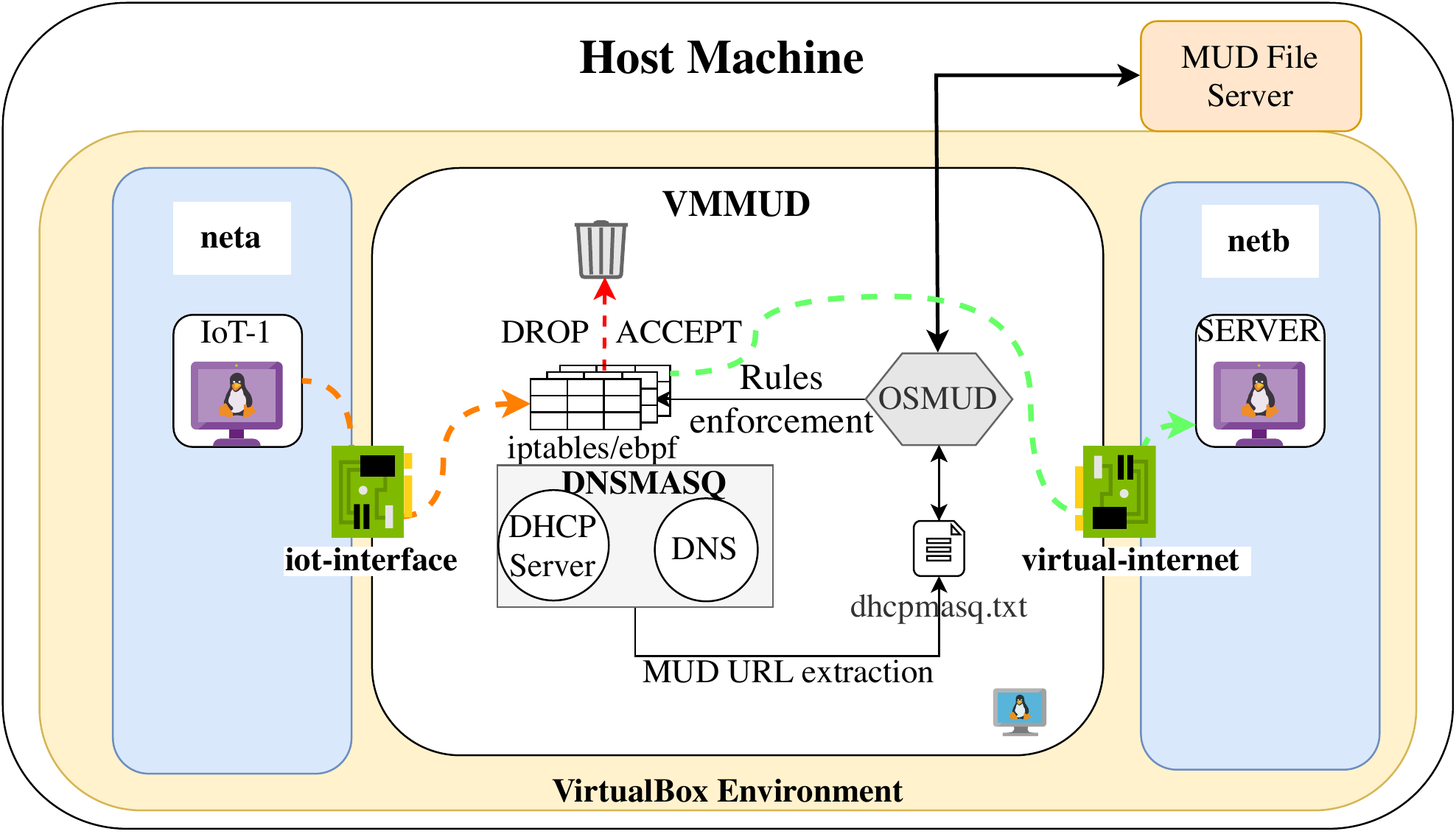}}
  \hfill
  \subfloat[Input flow\label{fig:networkdep-in}]{%
    \includegraphics[width=0.48\linewidth]{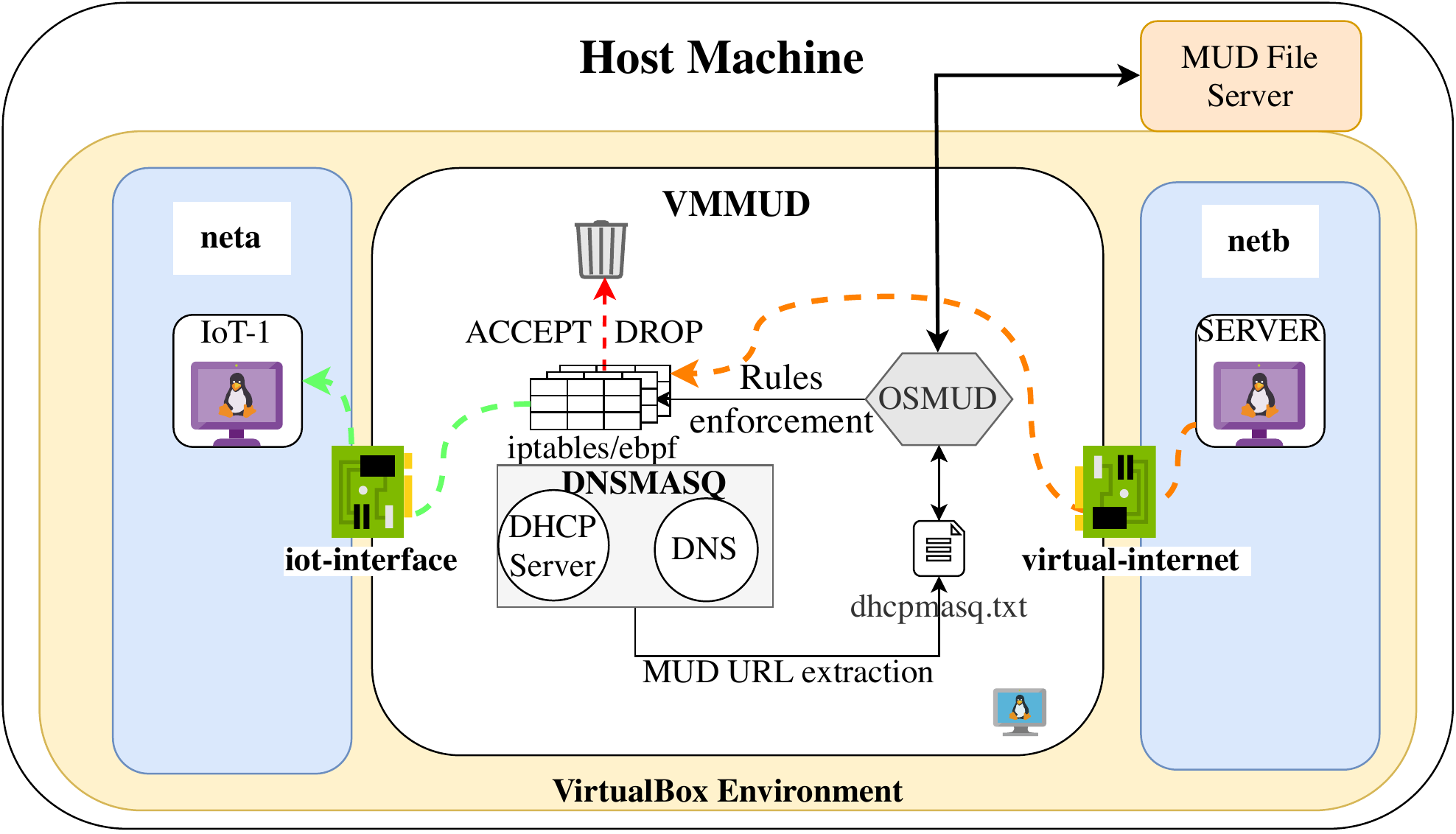}}
  \caption{Network deployment flows\label{fig:networkdep}}
\end{figure*}

Figure~\ref{fig:networkdep} shows the deployment of our prototype: a Linux Virtual Machine (VM) acting as a router and running an osMUD version using the extended MUD parser described in section \ref{ss:extending-mud-rate-limiting} (thereafter VMMUD), an external MUD file server and other VMs used for the evaluation. After installing on the VMMUD environment the libraries needed by osMUD to properly work, we built osMUD in its generic form. However, additional steps are required to deploy it correctly. Indeed, although the osMUD designers provided all necessary tools to build this version, there is still no support for a firewall other than OpenWRT, hence, the need to explore rule enforcement methods for commonly used firewalls. 

\begin{figure}
  \centering
  \includegraphics[width=\linewidth]{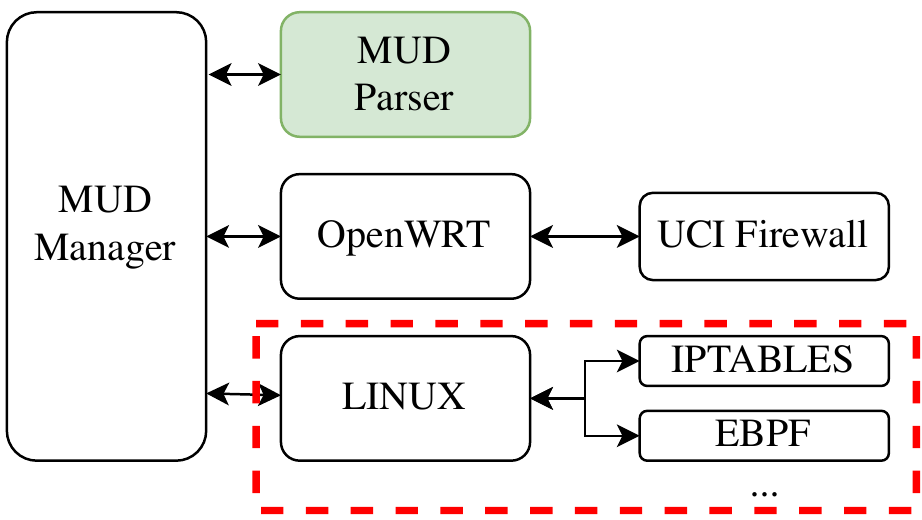}
  \caption{osMUD architectural blocks\label{fig:blocks}}
\end{figure}

To this end, the osMUD Manager structure allows developers to define additional rule enforcement methods by exploiting two main folders containing firewall integration code. The first includes several scripts that focus on rules enforcement and removal in an OpenWRT firewall, while the second folder is provided as a sample code modelling for new firewall integration. Hence, we consider these folders as containers of an \emph{adapter} that enables the MUD manager's independence from the underlying firewall (marked with red dotted line in Figure~\ref{fig:blocks}). We develop two new \emph{adapters} which are described in Section~\ref{s:adapting-mud}. The first one, \emph{eBPF-IoT-MUD} leverages eBPF for rule enforcement, and the second adapter leverages \emph{iptables}. Using either of these two new adapters, our system can be deployed on a standard Linux-based router or other constrained devices such as RaspberryPi.

To store and enforce rules produced from the new rate limit fields introduced in Section~\ref{s:extending-mud}, the osMUD manager parsing procedure needs to be updated accordingly. To achieve this, the MUD file parser (green block in Figure~\ref{fig:blocks}) analyses a new string of symbols in the group of action. Moreover, in order to properly store features enabling rate-limit operations, the parser uses an extended version of data types modelling MUD rules. Once the MUD manager receives these customised data types from the parser, it enforces the extended rules on the selected underlying firewall mechanism.

\section{Adapting MUD}
\label{s:adapting-mud}

MUD enforcement is carried out at the router using a backend that enables control over traffic. After introducing eBPF, XDP (eXpress Data Path) and \emph{tc}~\s{s:ebpf-xdp}, we describe in more detail the structure of an eBPF program~\s{s:ebpf-structure}, and then describe how we use eBPF and XDP in one backend implementation~\s{s:ebpf-adapter}. For comparison purposes, we also describe a second adapter that uses the Linux \emph{iptables} firewall support~\s{s:iptables}.

\subsection{eBPF, XDP, and \emph{tc}}
\label{s:ebpf-xdp}

The extended Berkeley Packet Filter (eBPF) is a set of instructions and a virtual machine (VM) for executing programs written in restricted C-language~\cite{lwnet, tutorial}. An eBPF program is ``attached'' to a specific code path in the kernel. When the code path is traversed, any attached eBPF programs are executed. They can be installed into the Linux kernel without modifying the kernel source code.

Thus, eBPF enables a variety of applications. For instance, an eBPF program can be attached to a network socket to perform tasks such as traffic classification or packet filtering. Furthermore, eBPF is useful for debugging the kernel and carrying out performance analysis, since eBPF programs can access kernel data structures.

The eXpress Data Path (XDP)~\cite{xdp} uses eBPF to enable fast packet processing at the lowest layer of the Linux network stack, immediately after a packet is received. XDP is the lowest layer of the Linux kernel network stack. It is present on the RX path, inside a device's network driver, meaning that it can process only the incoming packets. It allows packet processing at the earliest stage in the network stack, making it suitable for applications such as DDoS mitigation. The context received by an XDP program is defined by the type struct \texttt{xdp\_md}. The action returned by an XDP program is one of the following actions: the packet is dropped and raise an exception (\texttt{XDP\_ABORTED}), dropped silently (\texttt{XDP\_DROP}), passed along to the kernel stack (\texttt{XDP\_PASS}), retransmit on the same interface (\texttt{XDP\_TX}) or redirect to another target (for example, to another interface) (\texttt{XDP\_REDIRECT}).

Finally, the TC layer allows processing both egress traffic (transmitting packets) and ingress traffic (receiving packets). Traffic control policies on Linux are applied at this layer, with different queuing disciplines (\texttt{qdisc}) being configured for the different packet queues available in the system. Additionally, there is the possibility to add filters to drop or modify packets.

\begin{figure}
  \centering
  \includegraphics[width=\linewidth]{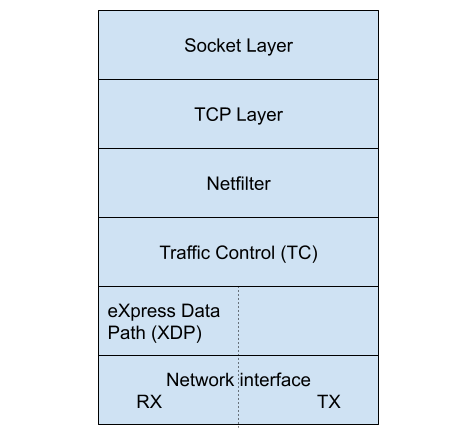}
  \caption{Linux kernel network stack showing XDP/TC~\cite{tutorial}
    \label{fig:kernel-stack}}
\end{figure}

\subsection{eBPF program structure}
\label{s:ebpf-structure}

eBPF programs can be loaded during runtime inside the Linux kernel, and they can interact with different kernel elements, such as kprobes, perf events, sockets and routing tables. An eBPF program can be attached to different \emph{network hooks}, eXpress Data Path (XDP) and Traffic Control (TC) (Figure~\ref{fig:kernel-stack}). It is executed whenever an event appears on the interface it is attached to. For example, in the case of an eBPF program that does custom packet processing, it will be executed whenever a packet is sent or received.

eBPF programs have a \emph{program type} which defines which layer or subsystem of the Linux kernel the eBPF program is attached to. The type provides information about: (i) what is the input passed to it (its context), (ii) which helper functions it is allowed to use, and (iii) to which kernel hook it will be attached to. For example, \texttt{BPF\_PROG\_TYPE\_SOCKET\_FILTER} is a program that does socket filtering, while \texttt{BPF\_PROG\_TYPE\_XDP} is program that is attached to the eXpress Data Path hook. A different category of programs are those for kernel tracing and monitoring.

eBPF \emph{maps} are key-value stores, where the keys and values can be user-defined data structures and types. Maps can be accessed from userspace as well as from eBPF programs loaded in the kernel, which makes them a powerful tool for communication between the two. Two common examples are \texttt{BPF\_MAP\_TYPE\_HASH} (which is similar to a hash table) and \texttt{BPF\_MAP\_TYPE\_ARRAY} (where entries are indexed by a number similar to an array). eBPF programs can use helper functions, such as functions for interacting with maps, for processing packet headers and others.

An eBPF program returns a \emph{code}, which depends on the program type. For example, an XDP program returns a code indicating what action regarding the packet after processing (pass packet, drop packet, redirect on another interface, retransmit on the interface it came from). Similarly, TC returns different codes (deliver the packet in the TC queue, drop packet, reclassify packet etc).

The eBPF in-kernel verifier performs a number of checks. The first check ensures that the eBPF program terminates and does not contain any loops. In the second check, the verifier simulates the execution of the eBPF program one instruction at a time to ensure that register and stack state are valid. Also, if pointer arithmetic is allowed, all pointer access are checked for type, alignment, and bounds violations. Uninitialised registers cannot be read. Certain registers are marked as unreadable, and checks are carried out to ensure that the read-only frame-pointer is not being written to. Lastly, the verifier restricts which kernel functions can be called from the eBPF programs, and which data structures can be accessed depending on their type.

\subsection{eBPF-IoT-MUD adapter}
\label{s:ebpf-adapter}

The osMUD manager can use the \emph{eBPF-IoT-MUD adapter} to enforce MUD rules in the lowest layer of the Linux kernel network stack. We implemented \emph{eBPF-IoT-MUD} as an XDP program, attached to the XDP hook.

There are two programs: \texttt{xdpfw\_from\_device} and \texttt{xdpfw\_to\_device}. The \texttt{xdpfw\_from\_device} is inserted on the LAN interface of the home router (Figure~\ref{fig:networkdep-out}). The \texttt{xdpfw\_to\_device} is inserted on the WAN interface of the home router (Figure~\ref{fig:networkdep-in}). The \texttt{xdpfw\_from\_device} is the program used to filter and/or rate-limit the connections made by the IoT devices to the Internet, while the \texttt{xdpfw\_to\_device} is the program used to filter and/or rate-limit the connections from the Internet to the IoT devices. In this paper, we are focusing only on \texttt{xdpfw\_from\_device}, as our aim is to stop DDoS attacks from IoT botnets. We note that \texttt{xdpfw\_to\_device} program functions in the same manner, enforcing MUD rules to prevent unauthorized access from the Internet to the IoT device in this case, thus minimizing the possibility of compromising an IoT device by an outside attacker~\cite{tax}.

We implemented a userspace program which attaches or detaches the programs from the specified network interface, and that can insert the appropriate MUD rules in \emph{allowlists}. There are two allowlists per protocol (\emph{IPv4 v4\_allowlist} and \emph{IPv6 v6\_allowlist}). The XDP programs use the allowlists to determine whether the packets will be allowed or dropped. The allowlists are implemented using eBPF maps (\texttt{BPF\_MAP\_TYPE\_HASH}).

\begin{figure}
  \begin{lstlisting}
    struct flow_key_ipv4 {
      __u8 ip_address_src[4];
      __u8 ip_address_dst[4];
      enum addr_type type;
      enum port_protocol proto;
      __u16 port;
    };
    struct flow_key_ipv6 {
      __u8 ip_address_src[16];
      __u8 ip_address_dst[16];
      enum addr_type type;
      enum port_protocol proto;
      __u16 port;
    };
    struct counters_rate {
      __u64 packets;
      __u64 bytes;
      __u64 max_pkt_rate;
      __u64 max_bytes_rate;
    };
  \end{lstlisting}
  \caption{eBPF-IoT-MUD maps\label{lst:ebpf-maps}}
\end{figure}

The key and value for the hash map can be seen in figure~\ref{lst:ebpf-maps}. The key in the hash map is represented by a structure \texttt{flow\_ipv4\_key} (and correspondingly \texttt{flow\_ipv6\_key}).

The key comprises the destination and source IP addresses for the \emph{from\_device} program, or the source and destination IP address for the \emph{to\_device} program, the destination port for \emph{from\_device}, or the source port for \emph{to\_device}, the protocol type and the type of rule (whether it is a \emph{from\_device} or a \emph{to\_device} rule). The type of rule is implicitly specified in the command line when specifying the type of port (source or destination port). The value in the hash map is represented by a structure \texttt{counters\_rate}, which records the number of bytes and the number of packets seen a rule. Additionally, the structure has a static maximum byte rate and maximum packet rate extracted from the extended MUD file model when the rule was created and inserted in the map. A default value of zero means that there is no rate-limit.

Whenever a packet arrives, the headers are parsed (layer 2, layer 3 and layer 4) and a key for the maps is built using the header information. The first header parsed is the Ethernet header, the second one is the IP header and, lastly, the transport header. The key is built using information from layer 3 and layer 4, according to the MUD standard. This key (\texttt{struct flow\_key\_ipv4} or \texttt{struct flow\_key\_ipv6}) is searched for in the allowlists. If the key is not found in the allowlists, the packet is dropped (using action \texttt{XDP\_DROP}). If it is found, the following actions will take place. First, the flow statistics for the matching entry (MUD rule) are updated according to the time window they fall in. If the current time window has expired, the statistics are first reset, and only then are updated. If the current time window has not expired, the flow statistics are updated directly. Next, the conditions for the maximum packet rate and/or maximum byte rate are checked. If these are not met (the current flow statistics for the current window are above the maximum thresholds), the packet is dropped (action \texttt{XDP\_DROP}), else it is passed along (action \texttt{XDP\_PASS}).

Moreover, an eBPF map of type \texttt{BPF\_MAP\_TYPE\_ARRAY} is used to store the current time to determine whether the packet falls in the current time window or whether the time window should be updated. Also, another eBPF map of type \texttt{BPF\_MAP\_TYPE\_HASH} is used to set the window size for updating statistics (by default one minute) from the userspace program.

\subsection{\emph{iptables} firewall adapter}
\label{s:iptables}

Referring to Figure~\ref{fig:blocks}, in this section we describe the \emph{iptables} \emph{adapter} that we implemented by analyzing the firewall integration with osMUD. The \emph{iptables} firewall relies on two main concepts: tables and chains, where tables are made of chains, while chains are made of rules. From our perspective, the rules included in a MUD file are enforced in the FORWARD chain, which handles filtering procedures on packets passing through the firewall.

Thus, after parsing a MUD file, the osMUD manager produces two types of rules: ACCEPT and DROP-ALL. The former represents the allowed communications described by device manufacturers, while the latter is added by the osMUD manager to blacklist all the other domains not described in the MUD file. We employ the custom chain concept provided by the \emph{iptables} tool to improve rules management. It allows us to define programs acting only on MUD-related policies, thus without interfering with those system-related.

For everyday IoT device categories, rate-limiting can further improve the protection against DoS attacks. To support these policies in the MUD model, we propose an extension allowing manufacturers to define device behaviours based on the traffic volume. Towards this end, our \emph{iptables} \emph{adapter} uses the \emph{hashlimit} module enabling the rate-limit match for a group of connections.

\section{Evaluation}
\label{s:evaluation}

We have extensively evaluated our system implementation by using the setup in Figure~\ref{fig:networkdep}. The IoT device and the osMUD manager each run in a separate virtual machine, hosted on a MacBook Pro Intel Core i5 with 8GB RAM. Both \emph{iptables} and \emph{eBPF-MUD-IoT} firewalls are configured on VMMUD, which acts as a bridge between two networks: \emph{neta} and \emph{netb}. The former contains an IoT device named IoT-1, i.e. VM that emulates a resource-constrained device, while the latter contains a server interacting with VMs of \emph{neta}.

\subsection{osMUD performance}

\begin{table}
  \caption{MUD Manager performance
    \label{table:performance}}
  \renewcommand{\arraystretch}{1.2}
  \begin{tabular}{ c c c c c l }
    \toprule
    Parsing & Enforcement & Total-1 file & Total-10 files & Total-50 files \\
    \midrule
    1 ms & 931 ms & 2.27 sec & 20.57 sec & 112.65 sec \\
    \bottomrule
  \end{tabular}
\end{table}

We expect that most of the MUD files parsing and rule enforcement to happen when the router is first initialised or rebooted, but IoT devices may be added later to the network. Table~\ref{table:performance} summarises basic performance measurements of our customised version of osMUD. The MUD file used for these measurements contains a single trusted host and produces four firewall policies. Lateral-movement rules are not supported by osMUD, as mentioned in Section~\ref{s:liberating-mud}, which led us to choose not to include them in the file used for testing. The fields \emph{Parsing} and \emph{Enforcement}, provided in the table, represent the amount of time needed by our prototype to process a MUD file based on our data model, and to create and enforce policies included in that file. These have been calculated as average values over 50 repetitions of the above measurements. Moreover, the table reports the time taken to retrieve, process, and enforce 1, 10, and 50 MUD files based on our model. It should be noted that, in the reported results, the frequency of DHCP request polling (5 seconds in osMUD) is not considered, and MUD file requests are forwarded to a local server, thus ensuring minimal communication delays. If the MUD file server is located in a different network accessed over the Internet this might introduce an additional delay depending on the network connection. 

\subsection{Rule management via eBPF}

\begin{table*}
  \caption{eBPF-MUD-IoT performance (time in nanoseconds)
    \label{table:ebpf-performance}}
  \renewcommand{\arraystretch}{1.2}
  \centering
  \begin{tabular}{ c c c c c c c c }
    \toprule
    Experiment & Min & Median & Avg & 90th & 99th & Max & Std.dev. \\
    \midrule
    Insert rule & 3740 & 4347.43 & 4533.43 & 5086.6 & 7143.62 & 15037 & 1019.61 \\
    Delete rule & 3426 & 3978 & 4576.58 & 4504.2 & 23551.4 & 49581 & 4145.33 \\
    Datapath & 140 & 249 & 556.15 & 1787.2 & 2978.95 & 24915 & 1089.96 \\
    \bottomrule
  \end{tabular}
\end{table*}

In this section, we present the results for micro-benchmarks we run on eBPF-IoT-MUD in Table~\ref{table:ebpf-performance}. While we expect that most of the MUD files parsing and rule enforcement to happen when the router is first initialised or rebooted, new IoT devices may become part of the IoT network dynamically. Thus, we evaluate the time it takes to insert and to delete rules in the eBPF maps, by inserting $255$ rules, and then deleting the $255$ rules inserted. For these experiments, we used a Lubuntu 20.04 virtual machine running in VirtualBox 6.1.14 on a laptop with an Intel quad-core i7 processor and 16GiB RAM.

\begin{figure*}
  \centering
  \subfloat[Rules insertion time in the IPv4 allowlist\label{fig:insert}]{%
    \includegraphics[width=0.33\linewidth]{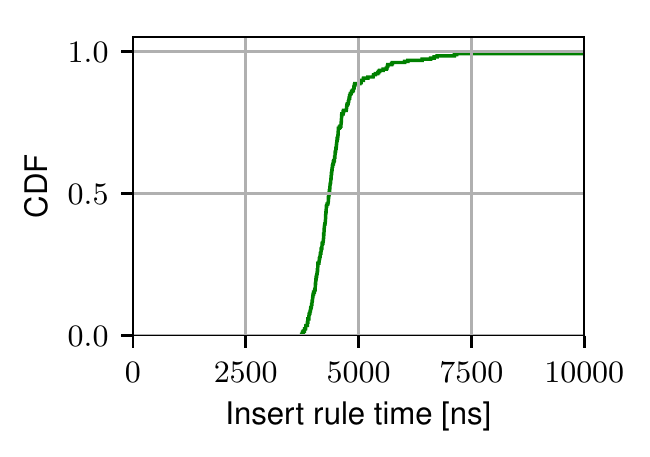}}
  \hfill
  \subfloat[Rules deletion time in the IPv4 allowlist\label{fig:delete}]{%
    \includegraphics[width=0.33\linewidth]{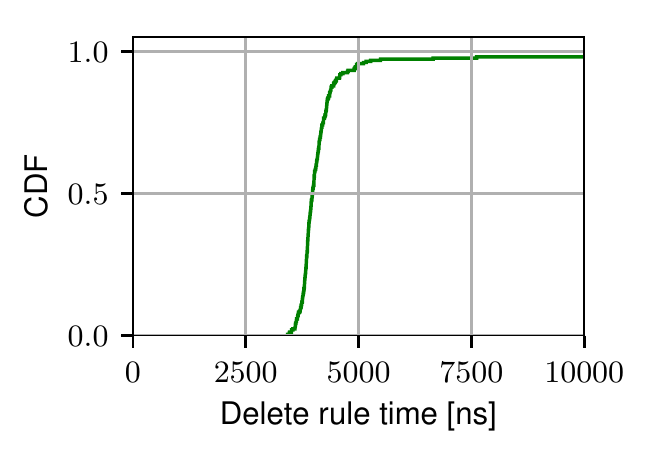}}
  \hfill
  \subfloat[Packet processing datapath\label{fig:match}]{%
    \includegraphics[width=0.33\linewidth]{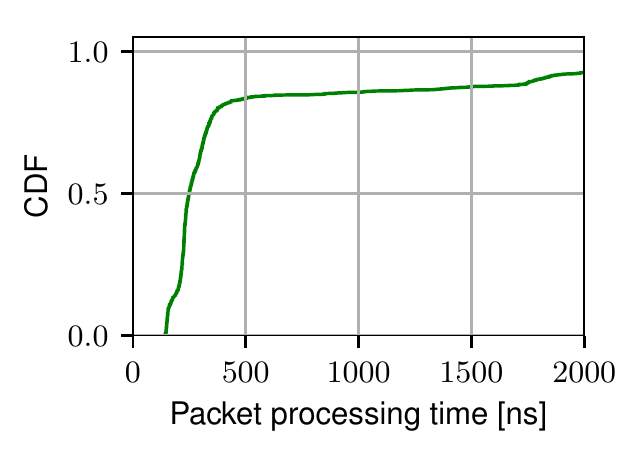}}
  \caption{eBPF-IoT-MUD evaluation}
\end{figure*}

Figure~\ref{fig:insert} presents the CDF for rules insertion time, with an average insertion time of 4533.43 ns, while Figure~\ref{fig:delete} presents the CDF for rules deletion time, with an average deletion time of 4576.58 ns. We also evaluated the packet processing datapath with one rule inserted for a ssh connection from outside the VM to the VM, and we sent different Linux commands on the ssh connection. Figure~\ref{fig:match} presents the CDF for of the packet processing datapath times, with an average of 556.15 ns.

\begin{table}
  \caption{Packet latency (time in microseconds) and transaction rate (transactions per second)
    \label{table:packet-latency}}
  \renewcommand{\arraystretch}{1.2}
  \centering
  \begin{tabular}{ l r r r r r }
    \toprule
    Experiment   & Min    & Avg    & Max    & Std.dev. & Txns/s \\
    \midrule
    No firewall  & 186.36 & 368.21 & 247286 & 1223.44  & 2714.01 \\
    iptables     & 187    & 345.5  & 203594 & 606.81   &  2892.51 \\
    eBPF-IoT-MUD & 188    & 342.72 & 258549 &  580.61  & 2915.91 \\
    \bottomrule
  \end{tabular}
\end{table}

We measure the packet latency using the experimental setup in Figure~\ref{fig:networkdep}, which runs on a laptop with an Intel quad-core i7 processor and 16 GiB RAM. The experiment runs \texttt{netperf}~\cite{netperf} with \texttt{TCP\_RR} on the client and sending traffic to the server across the router. On the router we inserted the MUD rules that allow the traffic to flow from the client to the server. We run netperf in three experiments: \one~the baseline (no firewall); \two~iptables firewall; \three~eBPF-IoT-MUD firewall, and we compare the packet latency and transaction rate.~\cite{netperf} runs five consecutive tests of 100 seconds for each experiment. The results are presented in Table~\ref{table:packet-latency}. There is no noticeable impact to packet latency and transaction rate compared to the baseline (no firewall) when using a firewall (iptables or eBPF-IoT-MUD), with the minimum packet latency being 186 us for baseline, 187 us with iptables and 188 us with eBPF-IoT-MUD. The transaction rate is 2714 transactions/s for baseline, 2892.51 transactions/s with iptables, and 2915.91 transactions/s with eBPF-IoT-MUD. These measurements show that using either firewalls does not add any additional overhead to packet processing.

\subsection{Rate limiting impact on normal traffic}

Based on the analysis we carried out in the previous section, we defined two MUD files, i.e., using "peaks" and "averages" policies, for each category considered, i.e. \emph{appliances} and \emph{smart-hubs}. We validate these MUD files using the dataset provided in~\cite{8440758} to emulate IoT device network traffic in normal conditions. The dataset includes traffic traces of $28$ different IoT devices over a period of $6$ months, of which only two weeks are openly available. We selected two devices for each category analysed, i.e. a Wi-Fi printer as \emph{appliance} and an Amazon Echo as \emph{smart-hub}. To make the traffic traces conform to our environment (Figure~\ref{fig:networkdep}), we used \texttt{tcprewrite} tool from the \texttt{tcpreplay} suite~\cite{tcpreplay}.

Once we processed these traces, we first generated and enforced MUD rules for \emph{iptables} firewall on the VMMUD machine using the osMUD manager. Secondly, we started the collection of iptables statistics in terms of total packets and bytes sent/dropped originating from the IoT device. Finally, to replay the devices' network traffic, we used \texttt{tcpreplay} on IoT-1 VM as shown in Figure~\ref{fig:networkdep-out}. We replayed three days of the TCP network traffic from each device selected for each MUD file defined, aiming to understand whether the policies generated allow the corresponding IoT device to function normally.

\begin{figure*}
  \subfloat[Average limit\label{fig:ntaverageLimit}]{%
  \includegraphics[width=.48\linewidth]{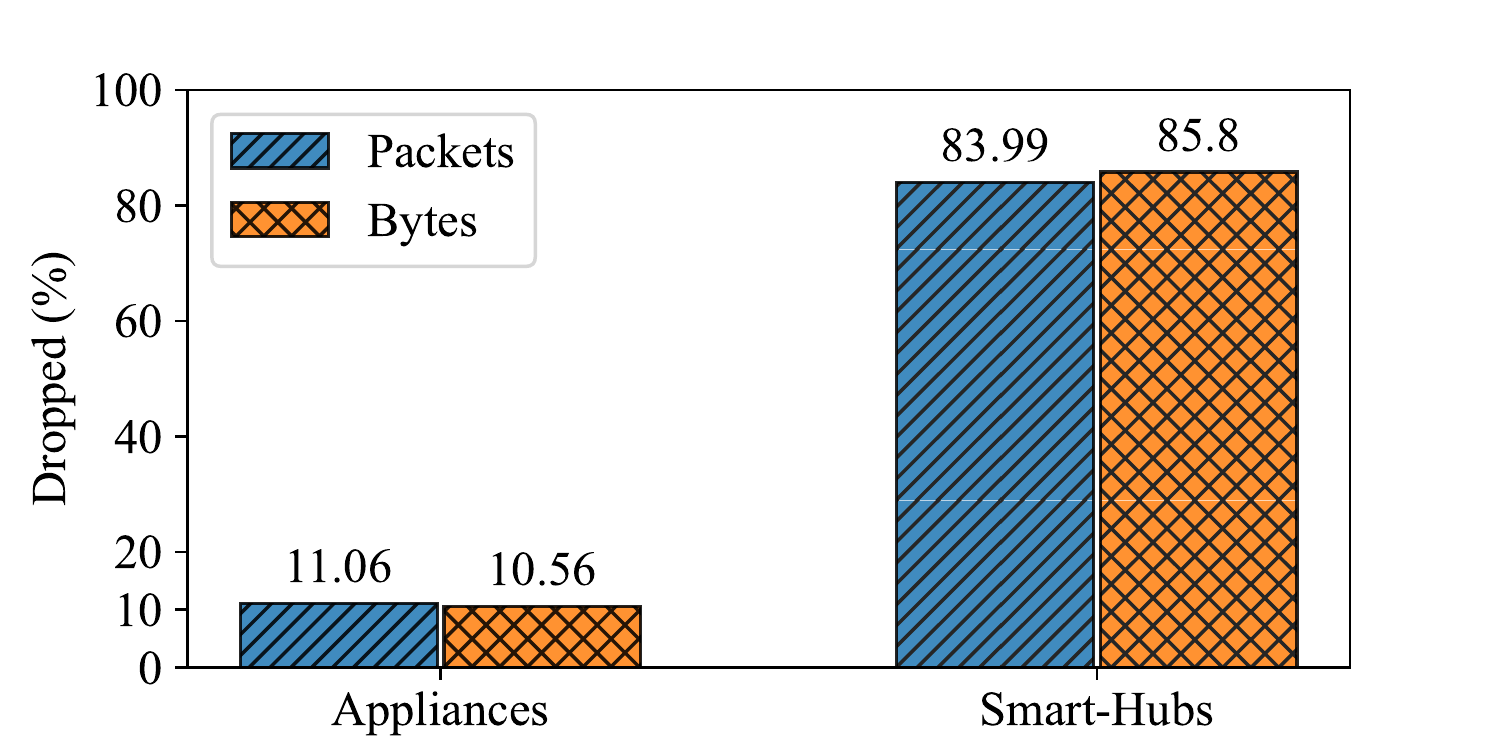}}
  \hfill
  \subfloat[Peak limit\label{fig:ntpeakLimit}]{%
    \includegraphics[width=.48\linewidth]{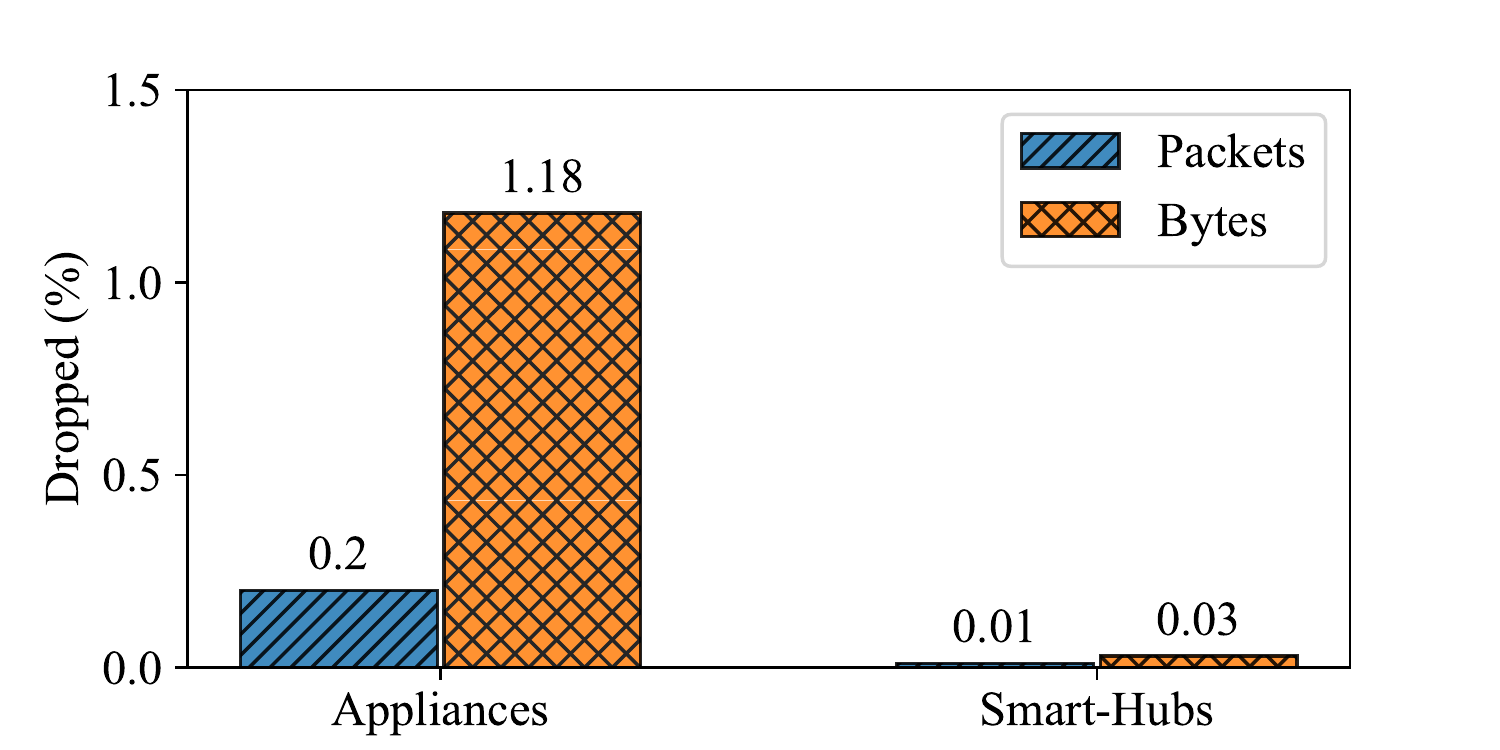}}
  \caption{Packet and byte drop rates under normal conditions
    \label{fig:normal-traffic}}
\end{figure*}

Figure~\ref{fig:normal-traffic} illustrates the percentage of packets and bytes dropped after applying these rate-limits. As shown in Figure~\ref{fig:ntaverageLimit}, the \emph{smart-hubs} MUD file using averages as rate-limits is hardly usable in normal device conditions, as it blocks most of the outgoing traffic (over 80\%). Similarly, average rate-limits affect the \emph{appliances} with a dropping rate of around 11\%, which might be a problem, especially during device updates. Conversely, as shown in figure~\ref{fig:ntpeakLimit} enforcing MUD rules using peaks as rate-limits does not affect the overall IoT traffic. 
In such a scenario, the packet drop rate remains below 1.5\% and 0.05\% in \emph{appliances} and \emph{smart-hubs}, respectively. Hence, to limit traffic volume for both categories, we decided to use MUD files defining peaks-based rate-limits. Furthermore, this choice is motivated by the fact that devices adopting TCP in output communications tend to reach the peak rapidly. On the one hand, it might be related to user activities, which may cause new iterations of the three-way handshake procedure. On the other hand, other main characteristics that make the TCP protocol reliable, e.g., packet re-transmission and congestion control, might represent the trigger of traffic peaks.

\subsection{Rate limiting impact on abnormal traffic}
\begin{figure*}
  \subfloat[Appliances rate-limit bytes per minute\label{fig:appliances-1}]{%
    \includegraphics[width=.48\linewidth]{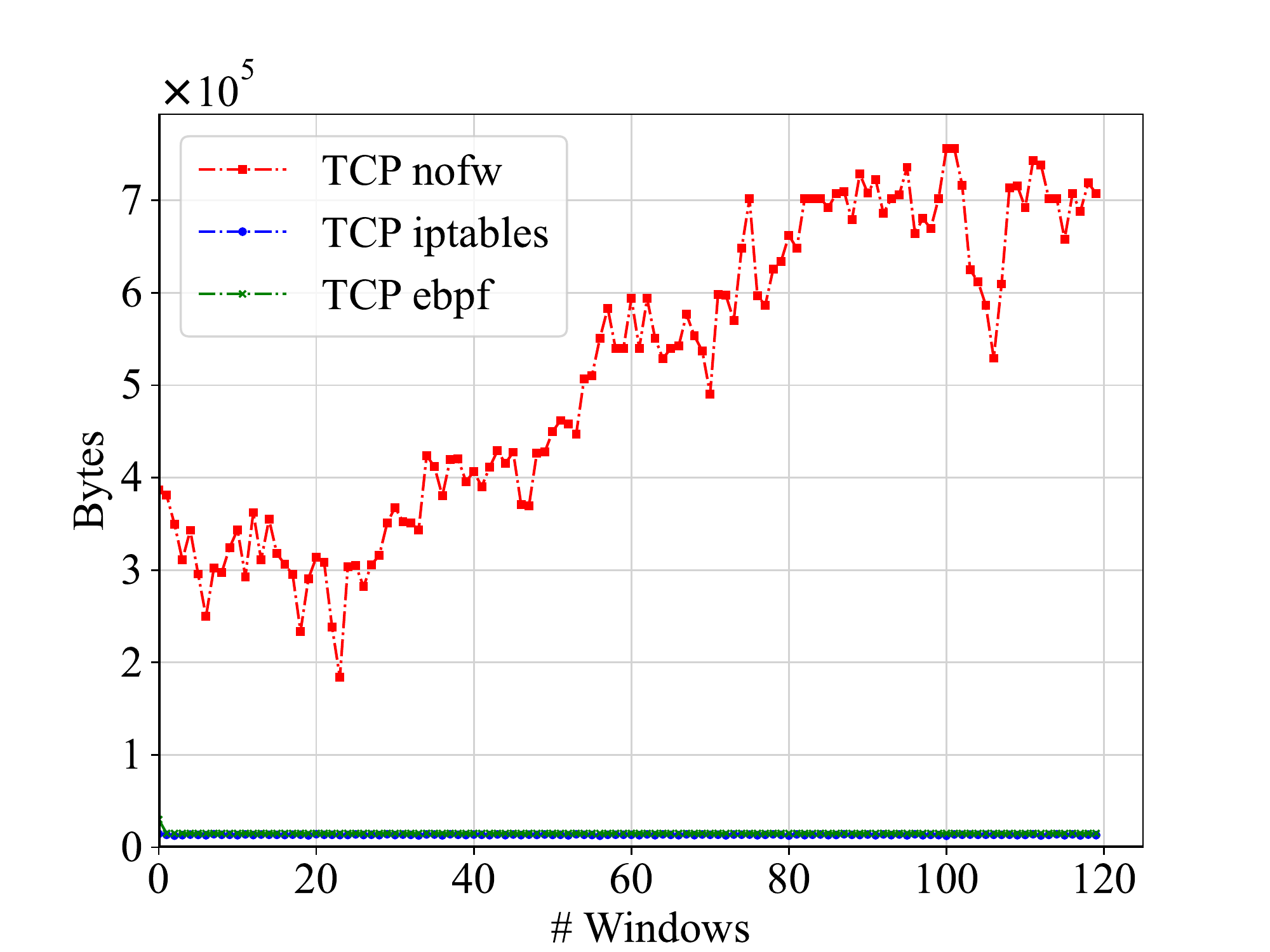}}
  \hfill
  \subfloat[Appliances rate-limit packets per minute\label{fig:appliances-2}]{%
    \includegraphics[width=.48\linewidth]{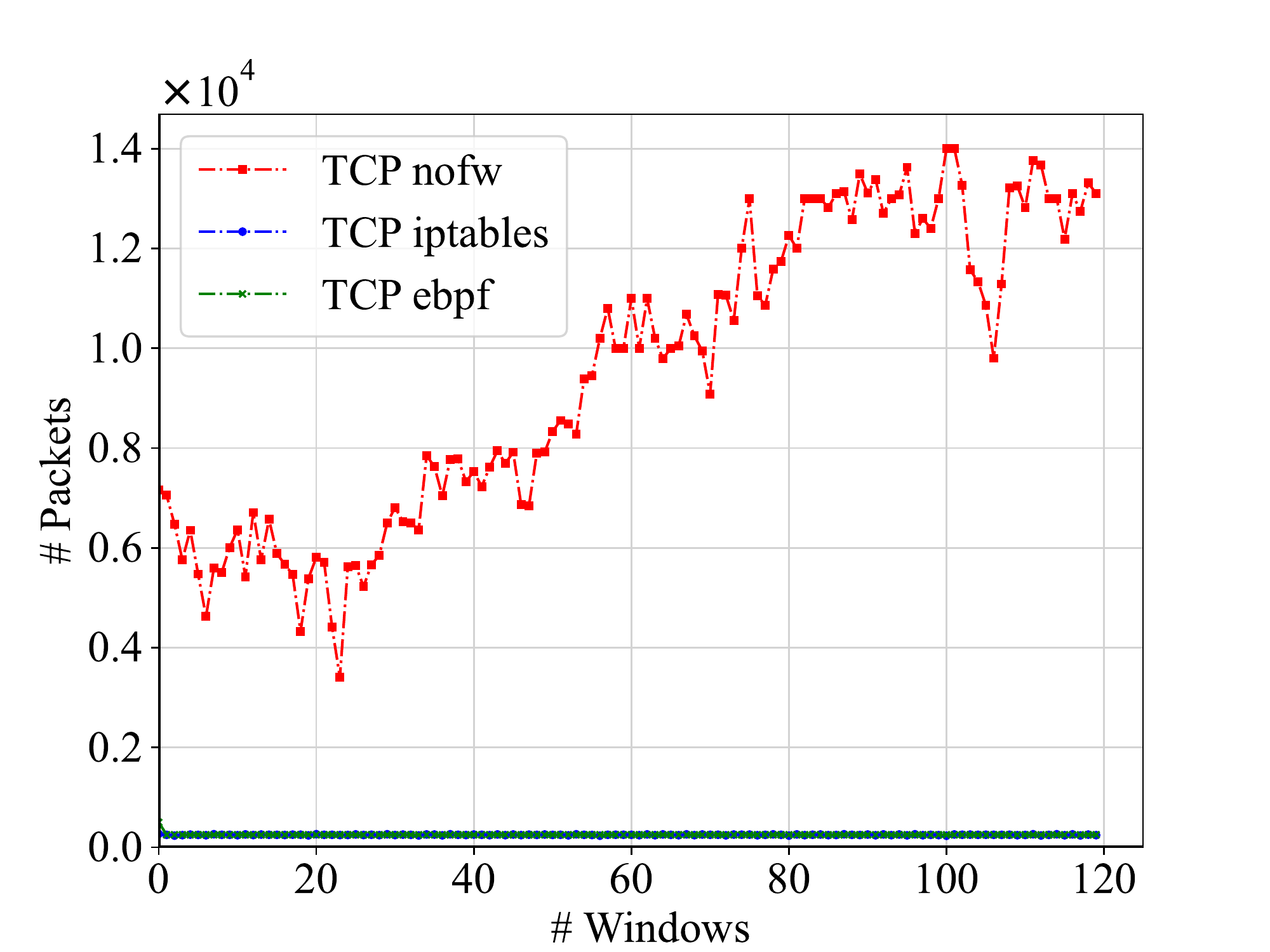}}
  \caption{Appliances  \label{fig:appliances}}
\end{figure*}

\begin{figure*}
  \subfloat[Smart hubs rate-limit bytes per minute\label{fig:smarthubs-1}]{%
    \includegraphics[width=.48\linewidth]{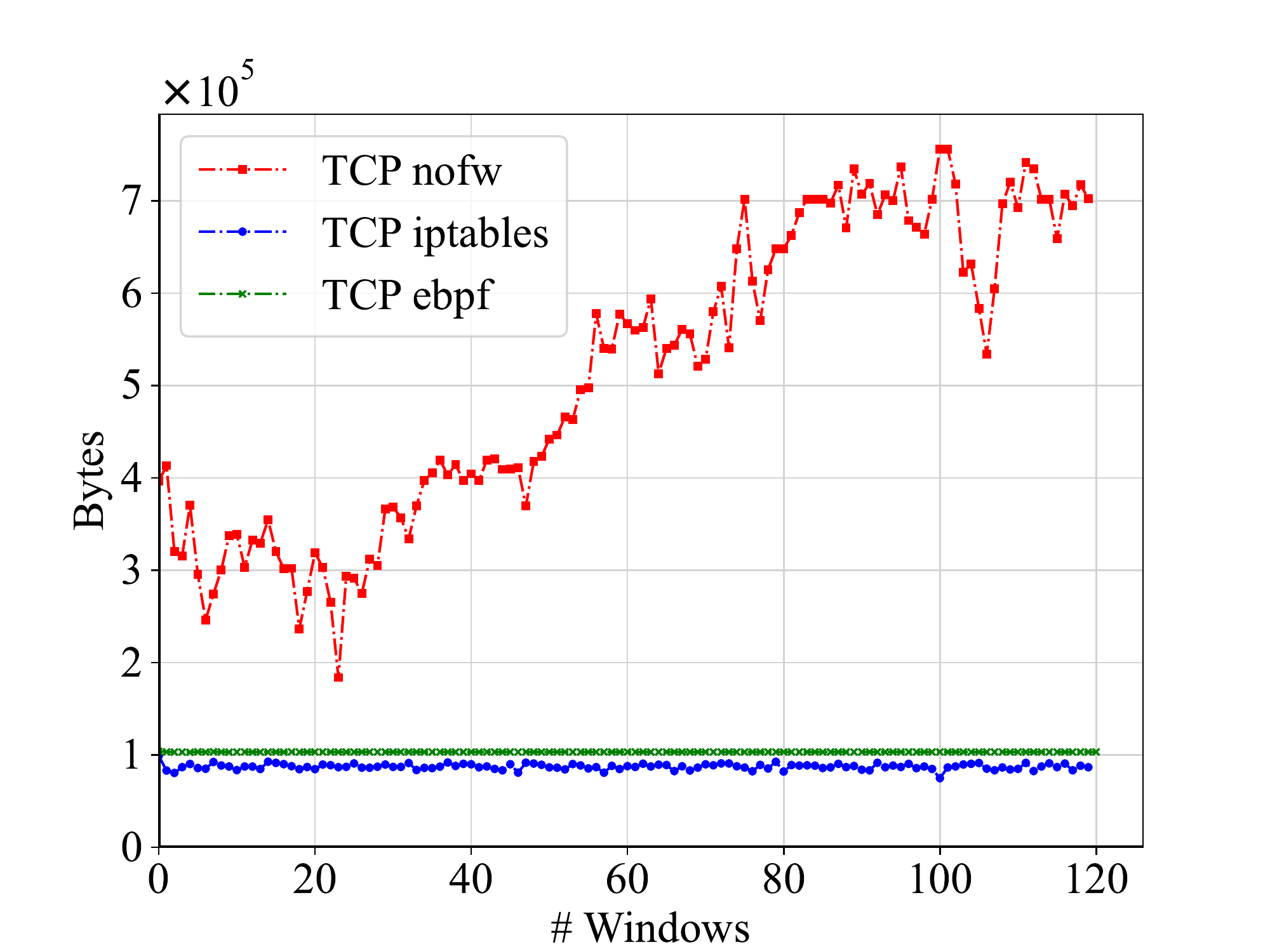}}
  \hfill
  \subfloat[Smart hubs rate-limit packets per minute\label{fig:smarthubs-2}]{%
    \includegraphics[width=.48\linewidth]{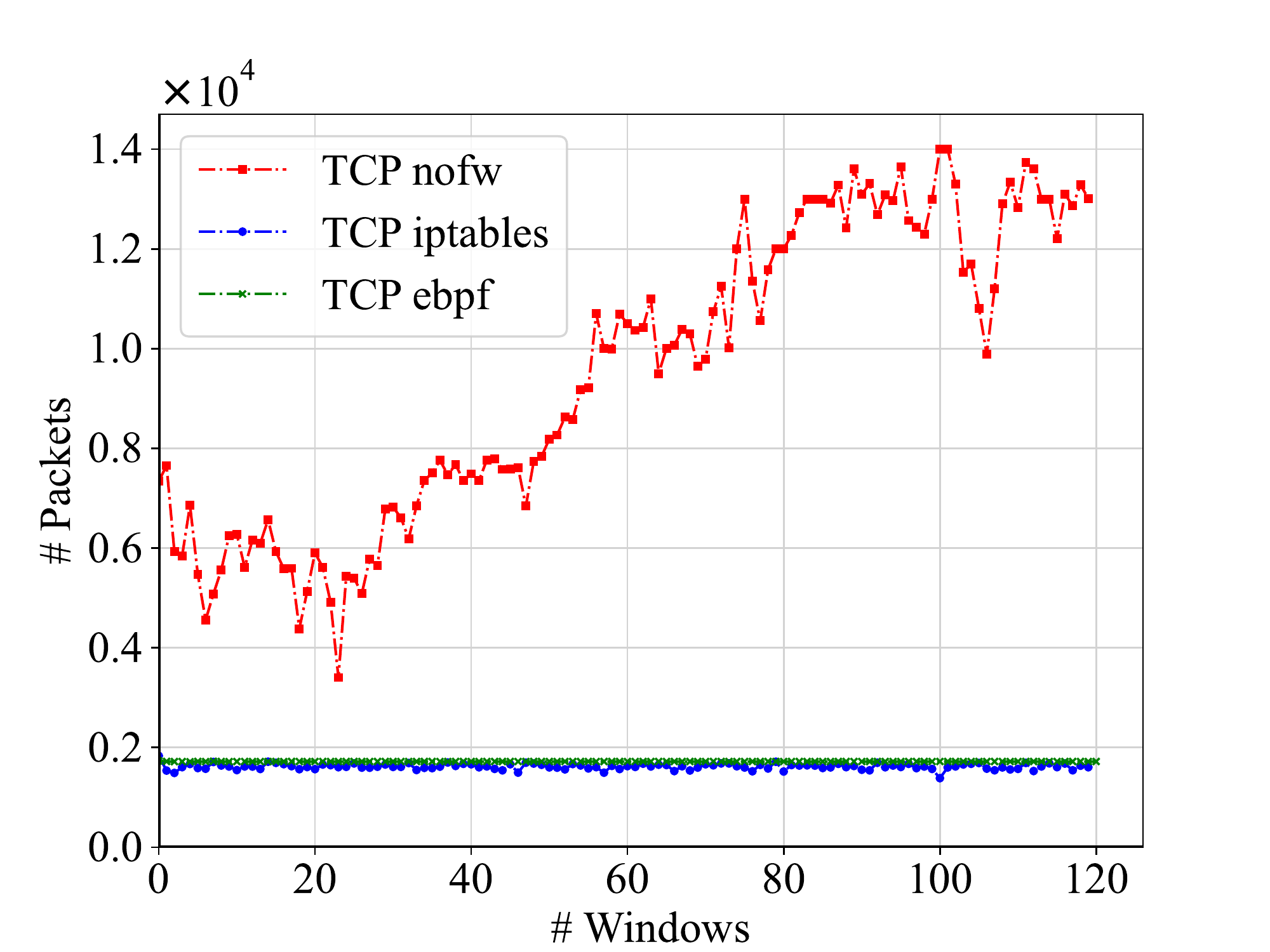}}
  \caption{Smart hubs
    \label{fig:smarthubs}}
\end{figure*}

To emulate abnormal IoT behaviors, we used the Network TON\_IoT dataset~\cite{ton-iot} providing network traces of several offensive systems conducting multiple attack scenarios, such as DoS, Ransomware, and injections attacks. We selected those referring to a DoS attack and merged them into a single trace. Next, we used the \texttt{tcprewrite} tool to rewrite the IP addresses in the trace to correspond to those of our test setup (Figure~\ref{fig:networkdep}). As described previously, to replay the traces we used \texttt{tcpreplay} on IoT-1, thus becoming  the originator of the SYN FLOOD attack, a form of DoS attack based on multiple SYN Request iterations. The traffic flow has been highlighted in Figure \ref{fig:networkdep-out}. Once we processed the abnormal network traces, we enforced the policies comprised in the MUD files previously selected. To test the effectiveness of the rate-limits against this attack, service ports do not appear in generated MUD files. 
Figure~\ref{fig:appliances} shows the results for the \emph{appliances} group while those \emph{smart hubs} related are illustrated in Figure~\ref{fig:smarthubs}. The red line shows the network traffic without rate-limiting, while the blue and green line represents the traffic after our shaping (using the rate-limits from the MUD files for the respective IoT device categories).

We run the experiment using the two firewall backends we implemented, based on iptables and eBPF-IoT-MUD. For iptables, we use an additional parameter, \texttt{--limit-burst}, that allows small traffic bursts (in our case bursts of additional $5$ packets). For eBPF-IoT-MUD, when the rate-limit is reached, the firewall starts dropping the incoming packets. This implementation difference explains why the traffic line in the Figure~\ref{fig:appliances} and Figure~\ref{fig:smarthubs} is straight when using eBPF-IoT-MUD, while when using iptables it is less smooth, but still meets the rate-limit with bursting allowed.

\subsection{Rate limiting discussion}
Although the results demonstrate the effectiveness of enforcing peak-based rules selected in our analysis, the scenario we presented assumes that the administrator - i.e., the user setting the MUD rules - may not have a comprehensive knowledge of the traffic flow generated in each communication that a device may engage in. In a more general case, where manufacturers define the MUD files, it becomes possible to enable a more precise approach to set rate limiting fields. For instance, a manufacturer can specify peak-based limits for every device they produce and for each outgoing communication (destination) from the device. This would help decrease the impact on device normal traffic and prevent DoS attacks from devices whose category aggregate rate limit is higher than their peak rate. We note that some DoS attacks might use traffic shaping, and thus might not be detected using a peak rate limit, in which case other botnet detection approaches are needed. 

One of our key contributions in this research is the inclusion of a rate limit field for a MUD rule. The value of these rate limits can be determined either by the users or the device manufacturers, although this is a separate research problem that falls outside the scope of our paper. Our focus, instead, is on demonstrating the effectiveness of enforcing these rate limits at the device level as a means of managing network interactions and bandwidth usage for IoT devices. By doing so, our proposed solution allows both knowledgeable users and manufacturers to mitigate DoS attacks that may target the services required by the device to operate normally, while minimizing the impact on the device's regular traffic. It is important to note that while our approach can be integrated with other solutions and techniques, it is not intended to be a comprehensive solution on its own, but rather a vital piece in the broader fight against DDoS attacks.
\section{Related Work}
\label{s:related}

IoT devices have been used in the past (such as the Mirai botnet~\cite{mirai}) to carry out Distributed Denial-of-Service (DDoS) attacks. The heterogeneity of IoT devices makes it difficult to manage these devices, while at the same time keeping the home network secure. IoT device non-clonable unique identity (the root of trust)~\cite{Nunes2021} is essential for authorizing the device on the network and setting a secure bidirectional connection using Public Key Infrastructure (PKI) or multi-factor authentication. However, considering the complexity of PKI solutions, and friction in user experience in multi-factor authentication, many IoT devices do not have these provisions~\cite{Diaz2019}, thus allowing devices to be vulnerable to many Botnet infested attacks~\cite{kolias2017ddos}.

Many IoT devices have one PKI for validating their own identities before setting outgoing traffic. However, there are no mechanisms for validating incoming traffic from other devices or the validity of other devices or users using their public keys. In addition to this, another set of impersonation attacks could happen even after the initial authentication has been set up successfully; for example, Bluetooth Impersonation Attacks (BIAs)~\cite{Antonioli2020} were recently detected in many Bluetooth IoT devices.

Similarly, other malware vulnerabilities recently detected in Wifi SDK led to more severe attacks through remote code execution~\cite{Claburn2021}. The exposure of these recent software vulnerabilities has provided a clear indication of a need for an additional layer of security. To add this extra layer of security, in this work, we explore the need of traffic monitoring and packet/byte rate filtering (limiting) at the router/hub/firewall level to prevent the damages created by victimised IoT devices ~\cite{Kim2021}. However, from the network management perspective, the configuration of filtering rules for each IoT device in the network is challenging due to the large variety of traffic patterns. To address this problem, the IETF RFC 8520~\cite{rfc8520} standard requires each IoT device service provider (or manufacturer) to provide a Manufacturer Usage Description (MUD) file profiling their IoT devices.

\paragraph{IoT and MUD}
Different organisations have developed prototypes of MUD manager (controller) implementations (middleware) in the last few years to allow seamless enforcement of the filtering rules specified in MUD files~\cite{osMud, nist_securing_2019}. Recent research surveys~\cite{andalibi2021analysis, Ramos2021, feraudo2020colearn} have discussed their limitations, challenges, and directions for future research ~\cite{bandwidth2020}.

One of the recognized challenges is the manufacturer's resistance to support MUD files and to make devices MUD-compliant in real deployment environments, in particular in absence of recognized IoT device usage patterns. One way to solve this problem is to create a behavioral fingerprint of the devices in the real environments and then use those fingerprints to design MUD files~\cite{Matthiasson2020}.

There have been significant efforts for creating behaviour fingerprints of IoT devices using their network traffic~\cite{Kolcun2021, Yadav2020}: the fingerprints are used to detect abnormal behaviours and take proper actions, either by limiting the incoming/outgoing traffic or by switching off the suspected anomalous devices. In the same direction, IoTrim~\cite{mandalari2021blocking}'s purpose is to limit the amount of communications that a device might have by denying those unnecessary for the primary and vital functions of the device in order to limit the information exposure of the user. PicP-MUD~\cite{picp-mud} analyses the content of the MUD flows to detect malicious traffic.

The enforcement of traffic filtering at the local system or router or hub level is done by introducing filtering rules through \textit{iptables}~\cite{Brown2019} or on SDN controllers~\cite{ranganathan2019soft, hamza2019detecting} or leveraging third-party online services to detect malicious destinations~\cite{habibi}. These works do not use eBPF and XDP to enforce traffic filtering. Furthermore, all of these works focus on traffic filtering based on source/destination address of the packets, and do not consider bandwidth or data rate. The idea of including bandwidth (packet rate and byte rate as a part of the MUD) has been discussed in new IETF RFC draft~\cite{bandwidth2020}. Andalibi et al.~\cite{Andalibi2019} discussed the advantage of peak request rate as a part of MUD when deploying MUD in fog computing environments. Similarly~\cite{Ramos2021, feraudo2020sok} have presented the challenges associated with the implementation of traffic rate control at the router and firewall levels in MUD-compliant networks. Currently, to the best of our knowledge, there is no MUD implementation that evaluates rate-limitation enforced by MUD in MUD-compliant networks. Our work shows and evaluates the advantages of including rate-limiting policies in MUD files, by showing with quantitative performance results that it is feasible to deploy extended MUD files using osMUD on Linux-based routers.

\paragraph{eBPF and XDP}
eBPF has emerged as a technology with applications in networking, tracing and profiling, security, and monitoring. XDP ensures efficient packet processing on the RX datapath, facilitating the construction of DDoS defences.~\cite{ebpf-itc} discusses the performance of packet filtering with eBPF.~\cite{miano} presents a hybrid system which uses XDP for traffic sampling and aggregation, and offloads DDoS mitigation rules to smartNICs.~\cite{ebpf-mon} uses eBPF and XDP for implementing traffic monitoring applications.~\cite{cloudflare} presents how XDP is integrated in the DDoS mitigation pipelines at Cloudflare to perform traffic analysis, aggregation, reaction and implement mitigation rules. Our work, eBPF-IoT-MUD, is the first to use eBPF in a smart home context, and implements a firewall using eBPF and XDP. Our custom system is integrated with the osMUD manager that provides it with the MUD rules to be enforced in order to prevent IoT devices from performing DDoS attacks on Internet destinations.

\paragraph{Learning to block}
IoTrim~\cite{mandalari2021blocking} finds the set of destinations contacted by an IoT device, and determines which of these destinations are needed to maintain device functionality. These destinations are usually related to the manufacturer, support or third-party analytics. If a destination is not needed for the device to operate normally, this destination will be blocked. The purpose of the work is to limit the information exposure of the IoT user because of privacy concerns. In our work, we present an integrated system that enforces MUD rules in order to avoid IoT devices contacting destinations that are not in the MUD file. Furthermore, based on the extended MUD file with rate limits, we thwart volumetric attacks such as DDoS attacks~\cite{volumetric2019} towards legitimate cloud destinations. Our work is orthogonal to IoTrim, since we do not make any judgements on whether the destinations in the MUD file our system is provided with are necessary for the IoT device to function normally, nor whether these destinations raise privacy concerns.

Our system can work in conjunction with IoT device identification systems that use machine learning models for device identification ~\cite{Kolcun2021}. As a first step, a manufacturer can provide a MUD file with rate limits already defined for an IoT device, or that has been built using a tool such as~\cite{clearasmud, mud-noms}. These rate limits can be further customized based on observed usage patterns using ML models for device identification. Furthermore, since smart homes can have different network connections and bandwidths, the network infrastructure conditions can further help to customize the MUD file.

\section{Conclusive Remarks}
\label{s:conclusions}
In the rapidly growing IoT ecosystem, the need for scalable deployment and management of IoT devices is becoming increasingly important. One of the significant challenges associated with IoT is the lack of security and privacy standards that can lead to serious security breaches and privacy violations. This is particularly challenging given the large number of IoT devices that are currently deployed and the complex interactions that occur between them. To address these challenges, the MUD specification was developed. The purpose of MUD is to restrict traffic while enabling IoT manufacturers to provide long-term support and updates for their devices, which can help to improve the security and privacy of IoT ecosystems. With MUD, IoT devices can be programmed to communicate only with approved services and networks, thereby preventing unauthorized access and protecting the privacy and security of the transmitted data.

This paper advances the related state-of-the-art along different directions. First, it originally presents the design of an extension to the MUD standard specification that rate-limits the outgoing traffic of IoT devices. This approach helps to ensure that IoT devices do not exceed their allocated bandwidth limits and do not negatively impact the performance of other devices on the network. We also demonstrate a procedure to identify rate-limits for consumer IoT devices. Secondly, we present the implementation and evaluation of a novel end-to-end system to enforce MUD rules on the home router through two new firewall backends, based on Linux standard iptables, and a custom system that uses eBPF (i.e., eBPF-IoT-MUD). Thirdly, we report novel experimental results about the performance evaluation of our MUD extension and system prototype by using the osMUD manager and by prototyping custom MUD rules in iptables and eBPF-IoT-MUD that are enforced at routers.



\ifCLASSOPTIONcaptionsoff
  \newpage
\fi



%

\bibliographystyle{IEEEtran}
\bibliography{references}

%








\end{document}